\documentclass[aps,prb,twocolumn,showpacs,preprintnumbers,amsmath,amssymb,
 4paper,superscriptaddress]{revtex4-1}
\usepackage{ifpdf}
 \newif\ifpdf
\ifx\pdfoutput\undefined
   \pdffalse
\else
   \pdfoutput=1
   \pdftrue
\fi
\ifpdf
   \usepackage{graphicx}
   \usepackage{epstopdf}
   \usepackage{color}
   \usepackage{verbatim}
\else
   \usepackage{graphicx}
\fi
\usepackage{bm}
\usepackage{srctex}
\usepackage{hyperref}

\DeclareMathOperator{\HKH}{\mathit{H}_{\mathrm{\scriptscriptstyle{KK}}}}
\DeclareMathOperator{\EFM}{\mathit{E}_{\mathrm{\scriptscriptstyle{FM}}}}
\DeclareMathOperator{\EAFM}{\mathit{E}_{\mathrm{\scriptscriptstyle{AFM}}}}
\DeclareMathOperator{\ENEM}{\mathit{E}_{\mathrm{\scriptscriptstyle{NEM}}}}
\DeclareMathOperator{\EDIM}{\mathit{E}_{\mathrm{\scriptscriptstyle{DIM}}}}

\begin{document}
\newcommand{\remark}[1] {\noindent\framebox{
\begin{minipage}{0.96\columnwidth}\textbf{\textit{ #1}}
\end{minipage}}
}

\newcommand{\question}[1] {\noindent\framebox{
\begin{minipage}{0.96\columnwidth}\textbf{\textit{Q: #1}}
\end{minipage}}
}

\newcommand{\bnabla}{{\boldsymbol{\nabla}}}
\newcommand{\eff}{\mathrm{eff}}

\def\red{\color{red}}
\def\blue{\color{blue}}
\def\green{\color{green}}

\title{ Magnetic phase diagram and quantum phase transitions in a two-species boson model}

\author{A.~M.~Belemuk}
\affiliation{Institute for High Pressure Physics, Russian Academy of Sciences, Moscow (Troitsk) 108840, Russia}
\affiliation{Department of Theoretical Physics, Moscow Institute of Physics and Technology  (State University), Moscow 141700, Russia}

\author{N.~M.~Chtchelkatchev}
\affiliation{Institute for High Pressure Physics, Russian Academy of Sciences, Moscow (Troitsk) 108840, Russia}
\affiliation{Department of Theoretical Physics, Moscow Institute of Physics and Technology  (State University), Moscow 141700, Russia}
\affiliation{L.D. Landau Institute for Theoretical Physics, Russian Academy of Sciences, Moscow  119334, Russia}
\affiliation{Institute of Metallurgy, {Ural Branch, Russian Academy of Sciences, Ekaterinburg 620016}, Russia}

\author{A.~V.~Mikheyenkov}
\affiliation{Institute for High Pressure Physics, Russian Academy of Sciences, Moscow (Troitsk) 108840, Russia}
\affiliation{Department of Theoretical Physics, Moscow Institute of Physics and Technology (State University), Moscow 141700, Russia}
\affiliation{National Research Center ``Kurchatov Institute'', Moscow 123182, Russia}

\author{K.I. Kugel}
\affiliation{Institute for Theoretical and Applied Electrodynamics, Russian Academy of Sciences, Moscow 125412, Russia}
\affiliation{National Research University Higher School of Economics, Moscow 101000, Russia}

\date{\today}

\begin{abstract}
We analyze the possible types of ordering in a boson--fermion model.  The Hamiltonian is inherently related to the Bose--Hubbard model for vector two-species bosons in optical lattices. We show that such model can be reduced to the Kugel--Khomskii type spin--pseudospin model, but in contrast to the usual version of the latter model, we are dealing here with the case of spin $S=1$ and pseudospin $1/2$. We show that  the interplay of spin and pseudospin degrees of freedom leads to a rather nontrivial magnetic phase diagram including the spin-nematic configurations.  Tuning the spin-channel interaction parameter $U_s$ gives rise  to quantum phase transitions. We find that the ground state of the system always has the pseudospin domain structure. On the other hand, the sign change of $U_s$ switches the spin arrangement of the ground state within domains from ferro- to aniferromagnetic one. Finally, we revisit the spin (pseudospin)-1/2 Kugel--Khomskii  model and see the inverse picture of phase transitions.
\end{abstract}


\maketitle

\section{Introduction}

The coherent manipulation (quantum simulation) of quantum systems  such as atoms in optical lattices, trapped ions, nuclear spins, superconducting circuits, or spins in semiconductors is currently a rapidly progressing field of physics. The experimental availability of quantum simulators is one of the recent scientific achievements~\cite{Lewens07_AP,hauke2012,George14_RoMP,Jurcevic2015PhysRevLett, Bermudez2017PRB,mazurenko2017Nature}. They allow one to simulate and explore a variety of complex many-body systems, in particular, the magnetic ones, and to verify the corresponding theoretical models yet unstudied or which can hardly (or even not at all) be directly investigated in solids~\cite{Greiner2002Nature,Kohl2005PRL,kawaguchi2012spinor,Flottat2017PhysRevB,Mondaini2017PhysRevB,Gao2017PhysRevLett}. The most actively studied quantum simulators are those implementing ultracold atoms in optical lattices~\cite{DeChi11_PRB,Gorshk10_NP,Porras04_PRL,Sun12_PRB,Sun14_PRB,Batrouni2009PhysRevLett}.  The interaction parameters characterizing such systems can be tuned \textit{in situ}.

In particular, the quantum simulators give an additional  impetus to the problem of boson magnetism that has attracted a growing interest  in recent years~\cite{Huerga2016PhysRevB,baier2016Science}  stimulated by novel experimental realizations in optical lattices~\cite{Lewenstein2015RepProgPhys,mazurenko2017Nature}.

Here, we consider a quantum simulator with  the strongly interacting Bose and Fermi degrees of freedom.   A possible realization is related to the Bose--Hubbard model for two species of vector bosons in optical lattices.

It is well known that the strongly correlated fermion Hubbard model, $H\sim t\sum_{<i,j>,\sigma} c_{i,\sigma}^+ c_{j\sigma}+U_0 \sum_i n_{i,\uparrow}n_{i,\downarrow}$, where $t$ and $U_0>0$ stand for the hopping and onsite Coulomb interaction,  respectively, reduces at half filling to the Heisenberg spin-$1/2$ model $H_{\eff}=J \sum_{<i,j>}\mathbf{S}_i\cdot \mathbf{S}_j$ with the antiferromagnetic exchange $J\sim  t^2/U_0>0$, $|t|\ll U_0$~\cite{Auerbach1994book}. Treating similarly the  strongly correlated Hubbard model for vector bosons, $H\sim t\sum_{<i,j>,\sigma} c_{i,\sigma}^+ c_{j\sigma}+U_0\sum_{i,\sigma} n_{i,\sigma}(1-n_{i,\sigma})$, one arrives at the Heisenberg spin model with ferromagnetic exchange $J\sim - t^2/U_0>0$~\cite{Imambekov03}.

In our case, we have two species of interacting vector bosons on the lattice that effectively leads to the appearing fermion degree of freedom, such as the  $1/2$-pseudospin. So, we have both: Bose and Fermi strongly interacting degrees of freedom. Naively, one can expect in the limit of strong correlation that this model reduces to some mixture of ferro- and antiferromagnetic interacting spins and pseudospins (p-spins). Nevertheless, the situation is more complicated. We show that the system can be described in terms of the  strongly anisotropic Kugel--Khomskii (KK) type spin--pseudospin model~\cite{Kugel1982UFN} with spin $S=1$ and pseudospin-$1/2$. Such type of the model has not been considered earlier in the context of optical lattices. Apart from the obvious case of optical lattices, the model in hand might have the physical realization in solid state (3$d$ metal compounds) for the case of KK model with p-spin 1 (triplet) and spin-1/2: one only has to interchange spin and p-spin to arrive to the model similar to that under consideration~\cite{Kugel1982UFN}.

We investigate below an interplay of spin and pseudospin  degrees of freedom. It is shown that tuning the spin channel interaction parameter $U_s$ leads to quantum phase transitions. We find that the ground state of the system always has the p-spin domain structure. On the other hand, the sign change of $U_s$ switches the spin arrangement of the ground state within domains from the ferro- to aniferromagnetic one. Finally, we revisit the (spin-1/2)--(p-spin-1/2) Kugel--Khomskii model and point out the dramatic difference between two types of the models: in particular we underline the inverse picture of phase transitions.

Our paper is organized as follows: In Sec.~\ref{sec:Hamilt}, we formulate the model and write down the Hamiltonian. In the next section, we derive the effective strongly anisotropic Kugel--Khomskii model with spin-1 and p-spin-1/2. In Sec.~\ref{sec:projHam}, we discuss the ground state of the Kugel--Khomskii model. Finally, Results and Discussions are presented in Sec.~\ref{sec:discuss}.

\section{Initial Model~\label{sec:Hamilt}}

\begin{figure}[t]
  \centering
  \includegraphics[width=0.4\columnwidth]{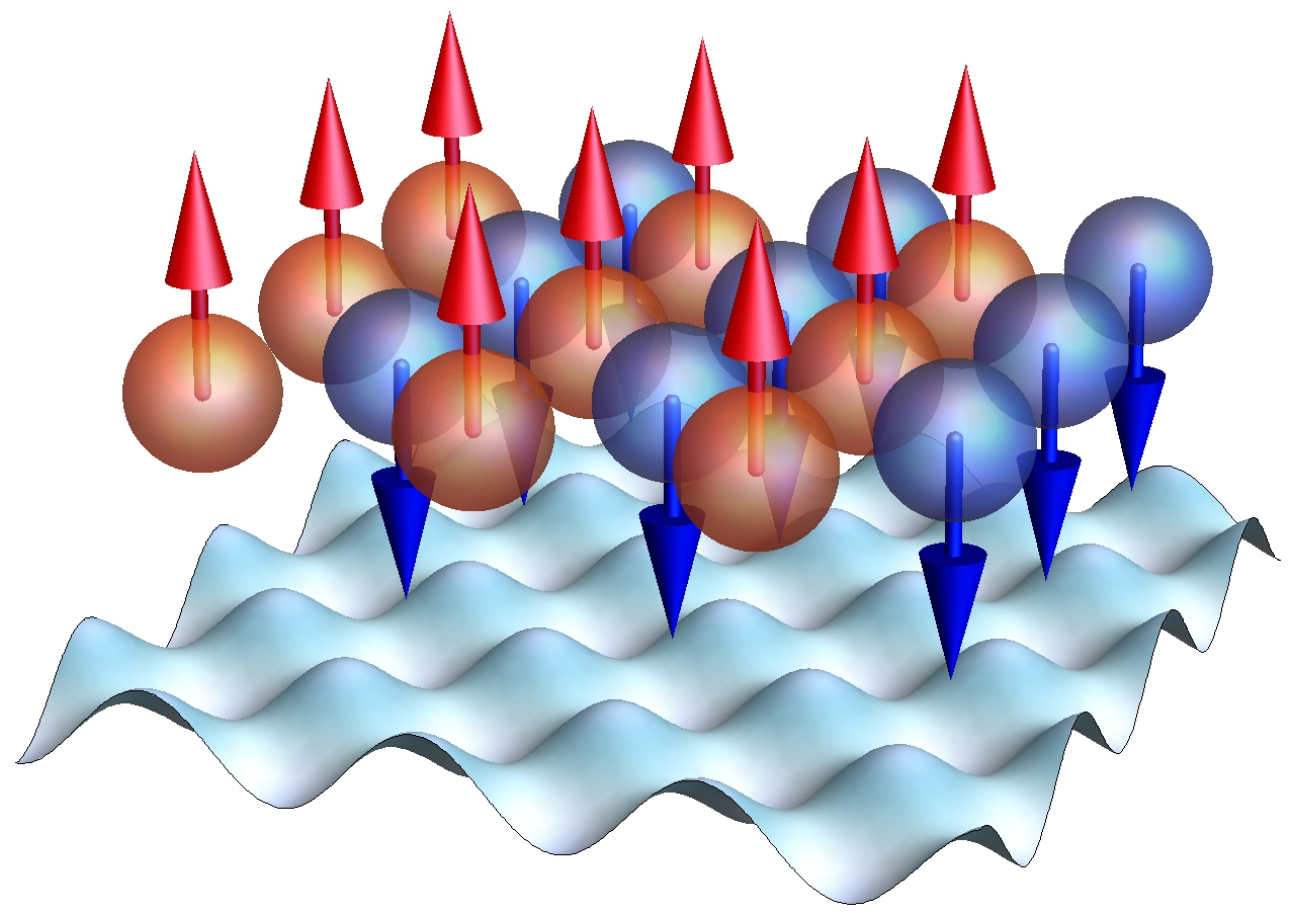}
  \caption{(Color online) Sketch of two types of vector bosons corresponding to cold atoms at the optical lattice.} \label{lattice}
\end{figure}

For two types of boson atoms with $S= 1$, we introduce the creation operators $ c^{\dagger}_{i \alpha s}$ for the states localized at site $i$ and having spin components $s= \{ -1, 0, 1 \}$. Index $\alpha= 1, 2$ accounts for two types of bosons. [In Fig.~\ref{lattice}, we show the example of atomic distribution over the lattice sites: the arrows correspond to spin and color denotes atom type.]

The total Hamiltonian is the sum of three terms:
\begin{gather}
 H=H^{\scriptscriptstyle (U_0)}+ H^{\scriptscriptstyle (U_s)}+ H_t.
\end{gather}
The interaction between bosons is given by two terms. The first one accounts for the Coulomb  repulsion between  two boson atoms occupying the same site~\cite{Lewenstein2015RepProgPhys}:
\begin{gather}
H^{\scriptscriptstyle (U_0)}=\sum_iU_{12} n_{i,1}  n_{i,2} +\sum_{i,\alpha=1,2} U_{\alpha\alpha}  n_{i,\alpha} ( n_{i,\alpha}- 1).
\end{gather}
Here, $U_{11}$, $U_{22}$, and $U_{12}$ are three interaction parameters playing the role of ``$U_0$'' discussed above and $ n_{i,\alpha}=\sum_s c^{\dagger}_{i \alpha s} c_{i \alpha s}$.  We assume below that bosons do not strongly differ from each other: $U_{12}\simeq U_{11}\simeq U_{22}=U_0$.

The spin-dependent interaction term, originating from the difference in scattering lengths for $S = 0$ and $S = 2$ total spin channels, is taken in the form (see, e.g., Refs.~\onlinecite{Imambekov03, Tsuchiya04})
\begin{gather}
 H^{\scriptscriptstyle (U_s)}_i= U_s( \mathbf S_i^2- 2n_i)/2.
\end{gather}
Here $n_i= n_{i,1}+ n_{i,2}$ is the total number of bosons at site $i$. The hopping term has the form:
\begin{gather} \label{hops}
  H_t= \sum_{\langle i,j \rangle} t_{\alpha} \left( c^{\dagger}_{i \alpha \sigma} c_{j \alpha \sigma}+ c^{\dagger}_{j \alpha \sigma} c_{i \alpha \sigma} \right),
\end{gather}
where the summation is performed over all nearest-neighbor sites  $\langle i,j \rangle$ and $\alpha=1,2$. The summation over the repeated indices is implied.

We show that two species of $S=1$ lattice bosons in the localized Mott insulating state belong to the well-known class of magnetic strongly anisotropic Kugel--Khomskii spin--p-spin models. This physical system is interesting due to the competition of several degrees of freedom:  particle density, spin and/or another degree of freedom (p-spin) distinguishing the boson species~\cite{anderlini2007Nature,folling2007Nature,trotzky2008Science,
Wagner2012PRA,Chtchelkatchev2014PRA}.

It is known for cold atoms that tuning interactions in the Mott insulators might generate new magnetic phases on top of spin or p-spin degrees of freedom~\cite{DeChi11_PRB,Gorshk10_NP,Porras04_PRL,Sun12_PRB,Sun14_PRB}.  Single species of spin $S=1$ lattice bosons have been already investigated~\cite{Imambekov03}. Here, we consider the two-species case, where the additional degree of freedom makes the problem more interesting.

 We show that tuning of the spin-channel interaction parameter $U_s$ leads to quantum phase transitions (QPT).  We find that the ground state of the system always has the p-spin domain structure. On the other hand, the sign change of $U_s$ switches the spin arrangement of the ground state within domains from ferro- to aniferromagnetic one.

\section{Strongly anisotropic Kugel--Khomskii model}

Hereinafter, we are focusing on the case where all the hopping  terms  are much smaller than all the interaction energies and the average site filling is unity. In the second-order perturbation theory in terms of $ t_{\alpha}$,  the effective Hamiltonian has the form $  H_{\eff}=  H_t (E_{\scriptscriptstyle G}-  H_0)^{-1}  H_t$. Here, $E_{\scriptscriptstyle G}= 0$ is the ground state energy of the Hamiltonian $ H_0=  H^{\scriptscriptstyle (U_0)}+ H^{\scriptscriptstyle (U_s)}$ for the average site filling $\langle n_i \rangle= 1$.

We can define spin $\mathbf S_i$ and p-spin $\bm{\mathcal T}_i$ operators~\cite{Yamashita1998PRB}:
\begin{gather}\label{eqSc}
  S^a_i= c^\dag_{i\alpha \sigma} s^a_{\sigma\sigma'}c_{i\alpha \sigma'},
  \qquad
  {\mathcal T}^a_i= c^\dag_{i\alpha \sigma} \tau^a_{\alpha\beta} c_{i\beta \sigma}.
\end{gather}
Below, we write down the resulting effective Hamiltonian in terms of these spin and p-spin operators.

In what follows, when we consider the link $\langle i, j \rangle$ between the nearest-neighbor sites, we focus on the basis of possible states for two bosons with spins $S_1= 1$ and $S_2= 1$ at neighboring sites $i=1$ and $j=2$. We are interested in the case with single occupation, i.e. when one boson of either type is located at each lattice site, $n_{i1}+ n_{i2}= 1$. For such case, there are four groups of states
\begin{gather}
\Phi^{1}_{ss'}= |ia\sigma {\,} ja\sigma' \rangle= a^{\dagger}_{i\sigma} a^{\dagger}_{j\sigma'} |0\rangle, \\
\Phi^{2}_{ss'}= |ia\sigma {\,} jb\sigma' \rangle= a^{\dagger}_{i\sigma} b^{\dagger}_{j\sigma'} |0\rangle, \\
\Phi^{3}_{ss'}= |ib\sigma {\,} ja\sigma' \rangle= b^{\dagger}_{i\sigma} a^{\dagger}_{j\sigma'} |0\rangle, \\
\Phi^{4}_{ss'}= |ib\sigma {\,} jb\sigma' \rangle= b^{\dagger}_{i\sigma} b^{\dagger}_{j\sigma'} |0\rangle,
\end{gather}
where for simplicity we explicitly distinguish operators for different types of bosons:  $ a_{i\sigma}\equiv c_{i,1,\sigma}$ and $ b_{i\sigma}\equiv c_{i,2,\sigma}$.

Each group contains $(2S_1+1)(2S_2+1)= 9$ states accounting for various possible spin components $s$ and $s'$ of  a boson located at site $i$ and another boson at site $j$. Combining the states with different $s, s'$ within one group, we can pass to the basis of the eigenstates of the total spin squared ${\bf S}^2= ({\bf S}_1+ {\bf S}_2)^2$ and its $z$-projection $S^z= S^z_1+ S^z_2$. We designate these states as $|S M \rangle$, $S= 0, 1, 2$ and $M=-S,\ldots,S$.
This basis can be written as follows
\begin{gather}
\Phi^{(1)}_{SM}=  |\phi^{(1)}_S \rangle |S M \rangle, \label{basis1}\\
\Phi^{(2)}_{SM}=  |\phi^{(2)} \rangle |S M \rangle, \\
\Phi^{(3)}_{SM}=  |\phi^{(3)} \rangle |S M \rangle, \\
\Phi^{(4)}_{SM}=  |\phi^{(4)}_S \rangle |S M \rangle. \label{basis4}
\end{gather}

The orbital part of wave functions for two identical ``$a$'' bosons, $|\phi^{(1)}_S \rangle$, or ``$b$'' bosons, $|\phi^{(4)}_S \rangle$, depends on the value of the total spin $S$.  In particular, for two $a$ bosons, we have (remind, that the total boson wave function, incorporating spin and p-spin degrees of freedom, has the proper bosonic symmetry):
\begin{gather}
|\phi^{(1)}_S \rangle = \frac{1}{\sqrt{2}} \bigl( |ia {\,} ja \rangle + |ja {\,} ia \rangle \bigr), \quad S= 0, 2, \\
|\phi^{(1)}_S \rangle = \frac{1}{\sqrt{2}} \bigl( |ia {\,} ja \rangle - |ja {\,} ia \rangle \bigr), \quad S= 1,
\end{gather}
and for two $b$-bosons
\begin{gather}
|\phi^{(4)}_S \rangle = \frac{1}{\sqrt{2}} \bigl( |ib {\,} jb \rangle + |jb {\,} ib \rangle \bigr), \quad S= 0, 2, \\
|\phi^{(4)}_S \rangle = \frac{1}{\sqrt{2}} \bigl( |ib {\,} jb \rangle - |jb {\,} ib \rangle \bigr), \quad S= 1.
\end{gather}

The orbital functions for two bosons of different types do not depend on the total spin $S$ and have the form
\begin{gather}
|\phi^{(2)} \rangle = |ia {\,} jb \rangle= \phi_i({\bf r}_1)\psi_j({\bf r}_2), \\
|\phi^{(3)} \rangle = |ib {\,} ja \rangle= \psi_i({\bf r}_1)\phi_j({\bf r}_2),
\end{gather}

The p-spin operator in Eq.~\eqref{eqSc} is responsible for transitions between different types of bosons. We express $H_{\eff}$ in terms of  $\mathbf S_i$ and  $\bm{\mathcal T}_i$.  Since we assume that bosons do not strongly differ from each other, then  $t_1\simeq t_2=t$. Note that this condition does not mean the absence of any difference between two types of bosons.

This difference manifests itself in the absence of  the cross-terms $\sim c^{\dagger}_{i 1 \sigma} c_{j 2 \sigma}$  in the hopping Hamiltonian \eqref{hops} that allows tunneling with the interchange of boson type (with initial state with one sort of boson and the final state --  with another sort of boson). Formally, it means the conservation of the p-spin projection in the course of tunneling.

\begin{widetext}
Finally, we arrive at the magnetic model of the Kugel--Khomskii type~\cite{Kugel1975FTT,Kugel1998PRB,Khomskii2003JMMM}
\begin{gather}\label{eq:Heff}
 H_{\eff}= \sum \limits_{\langle i,j \rangle} \left\{  \left(\epsilon +  J {\bf S}_i \cdot {\bf S}_j+ K  ({\bf S}_i \cdot {\bf S}_j)^2 \right) \left[\frac{1}{2}+ 2 {\mathcal T}^z_i {\mathcal T}^z_j\right]+
\left( \epsilon' +  J' {\bf S}_i \cdot {\bf S}_j+ K' ({\bf S}_i \cdot {\bf S}_j)^2 \right) \left[ \frac14- 2 {\mathcal T}^z_i {\mathcal T}^z_j+ {\bm {\mathcal T}}_i \cdot {\bm {\mathcal T}}_j \right] \right\}.
\end{gather}
\end{widetext}

Effective exchange integrals are
\begin{gather}
J=-\frac{1}{(1+\lambda)},\qquad K=-\frac1{(1-2\lambda)(1+\lambda)},
\\
J'= \frac{ 2\lambda}{1- \lambda^2}, \qquad K'=-\frac{2\lambda^2}{(1- \lambda^2)(1- 2\lambda)},
\end{gather}
and
\begin{gather}
\epsilon=\frac{2 \lambda}{(1+\lambda)(1- 2\lambda)},
\\
 \epsilon'=-\frac{2 (1- \lambda^2- 2\lambda)}{(1-\lambda^2)(1- 2\lambda)},
\end{gather}
where  $\lambda= U_s/U_0$ and the unit of energy is $E_u= 2t^2/U_0$. Note that $H_{\eff}$ involves the projection operators in the p-spin subspace:
\begin{gather}\label{eqP}
 P= \frac{1}{2}+ 2 {\mathcal T}^z_i {\mathcal T}^z_j
 \\\label{eqP'}
 P'=\frac14- 2 {\mathcal T}^z_i {\mathcal T}^z_j+ {\bm {\mathcal T}}_i \cdot {\bm {\mathcal T}}_j,
\end{gather}
 where the first one projects onto the state $|T= 1, M_T= \pm 1 \rangle$ and the second one -- onto the state $|T= 1, M_T= 0 \rangle$. Here, $T$ is the total p-spin of the link and $M_T$ is its projection.  Note also that parameter $\lambda$ plays here the role similar to that of the ratio of the Hund's rule coupling constant and the on-site Coulomb repulsion in usual compounds with the orbital degeneracy~\cite{Kugel1975FTT,Kugel1998PRB,Khomskii2003JMMM}.
\begin{figure}[tb]
  \centering
  \includegraphics[width=0.99\columnwidth]{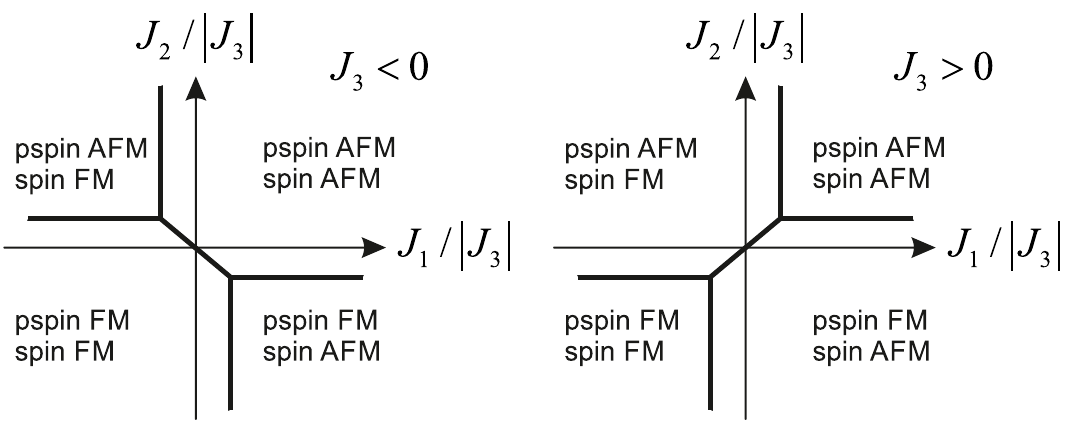}
  \caption{ A sketch of the phase diagram corresponding to the simplest $S=1/2$ and $\mathcal T=1/2$ symmetrical Kugel--Khomskii Hamiltonian $ H_{\rm sym}= \sum_{\langle i j \rangle}\{J_1\,{\mathbf S}_i\cdot {\mathbf S}_j+J_2\,{\bm {\mathcal T}}_i \cdot {\bm {\mathcal T}}_j +4J_3({\mathbf S}_i\cdot {\mathbf S}_j)\,({\bm {\mathcal T}}_i \cdot {\bm {\mathcal T}}_j)\}$, in the mean-field approximation~\cite{Kugel1998PRB}. }\label{figkhphase}
\end{figure}

\section{Projected Hamiltonian~\label{sec:projHam}}

The usual way to understand the main features of the complicated phase diagram of the Kugel--Khomskii type model is the mean-field approach~\cite{Kugel1998PRB}. Apart from the low dimensional systems it usually captures the main physical features of the phase diagram. Indeed, even an evident overestimation of the role of quantum fluctuations in the two-site solution of the model Hamiltonian reported in Ref.~\onlinecite{Kugel1998PRB} does not affect qualitatively the form of the phase diagram shown in Fig.~\ref{figkhphase}.  Below we follow the mean-field approach dealing with  Hamiltonian~\eqref{eq:Heff}.

Within the mean-field approach, the ground state of the Kugel--Khomskii type Hamiltonian usually reduces to AFM (FM) spin and p-spin arrangements like it is sketched in the phase diagram for the simplest $S=1/2$ and $\mathcal T=1/2$ symmetrical Kugel--Khomskii model Fig.~\ref{figkhphase}. [We remind that in our case the Kugel--Khomskii type model is strongly asymmetrical and $S=1$.] For $H_{\eff}$, this is also true as we have found. So, there are two basic types of p-spin arrangements.

The first one  arises when the lattice is filled with one type of bosons or when atoms of one sort bunch into domains on the lattice. Then, we have ferromagnetic p-spin wave function 
\begin{gather}\label{equpup}
 |\Uparrow\rangle= \prod _i|+\rangle_i,
\end{gather}
(or $ |\Downarrow\rangle= \prod _i|-\rangle_i$).  Here ``+ (-)'' stand for p-spin up (down). 
 
 The second type of p-spin arrangement takes place when two types of bosons are alternating  at the neighboring sites. The p-spin wave function is of the antiferromagnetic type 
 \begin{gather}\label{equpd}
 |\upharpoonleft\!\downharpoonright\rangle= |+ - + - \dots \rangle.
\end{gather}

\begin{figure}[tb]
  \centering
  \includegraphics[width=0.99\columnwidth]{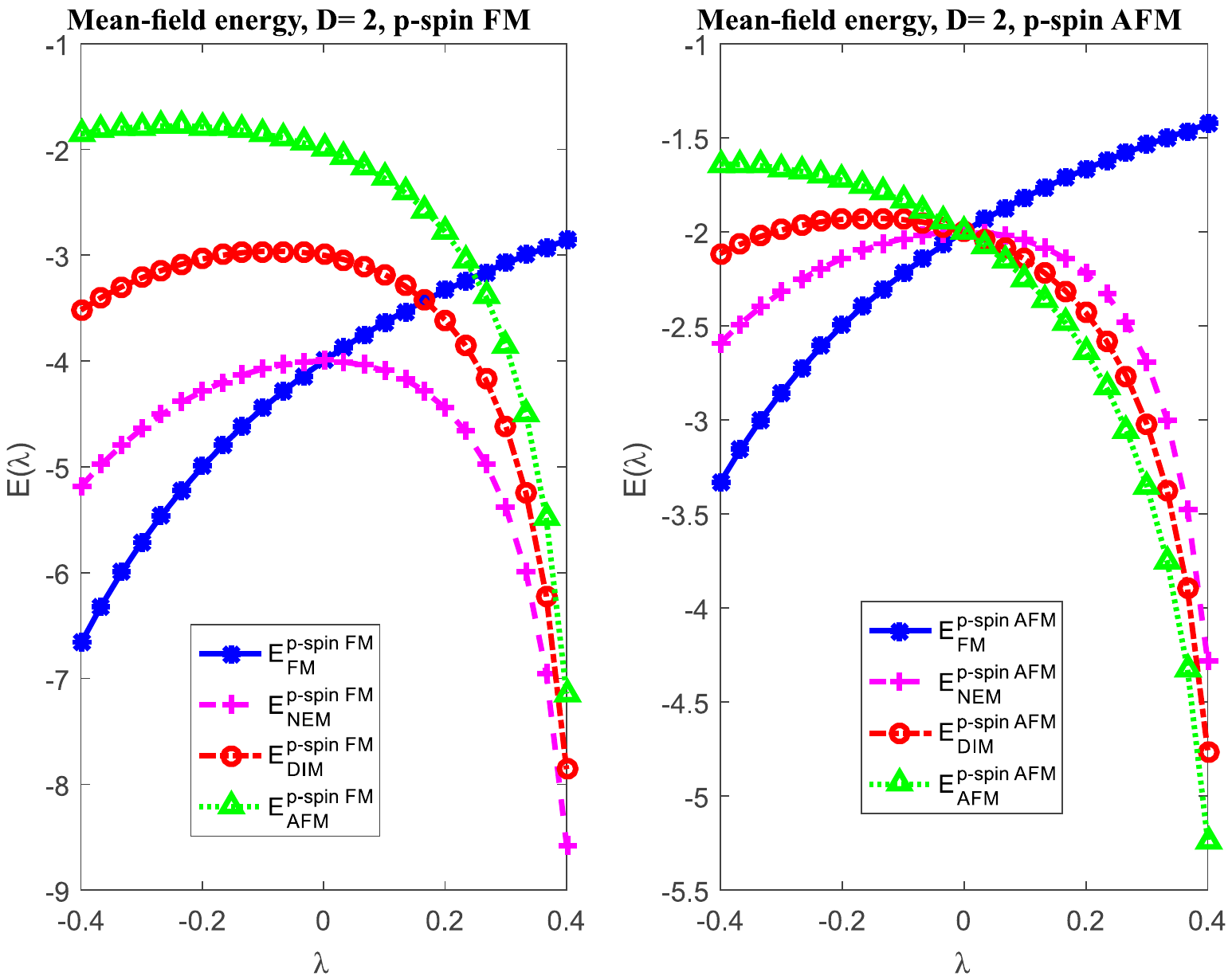}\\ 
  \caption{(Color online) Mean-field energies of $H_{\eff}$ corresponding to two kinds of $S=1$ bosons on the 2D square optical lattice for the average site filling $\langle n_i \rangle= 1$ (the total number of bosons at site $i$). The energy is normalized by $E_u=2t^2/U_0$.   Left panel is for FM p-spin arrangement, right panel --- for AFM p-spin arrangement. }\label{FigVectorBosons}
\end{figure}
\begin{figure}[t]
  \centering
  \includegraphics[width=0.99\columnwidth]{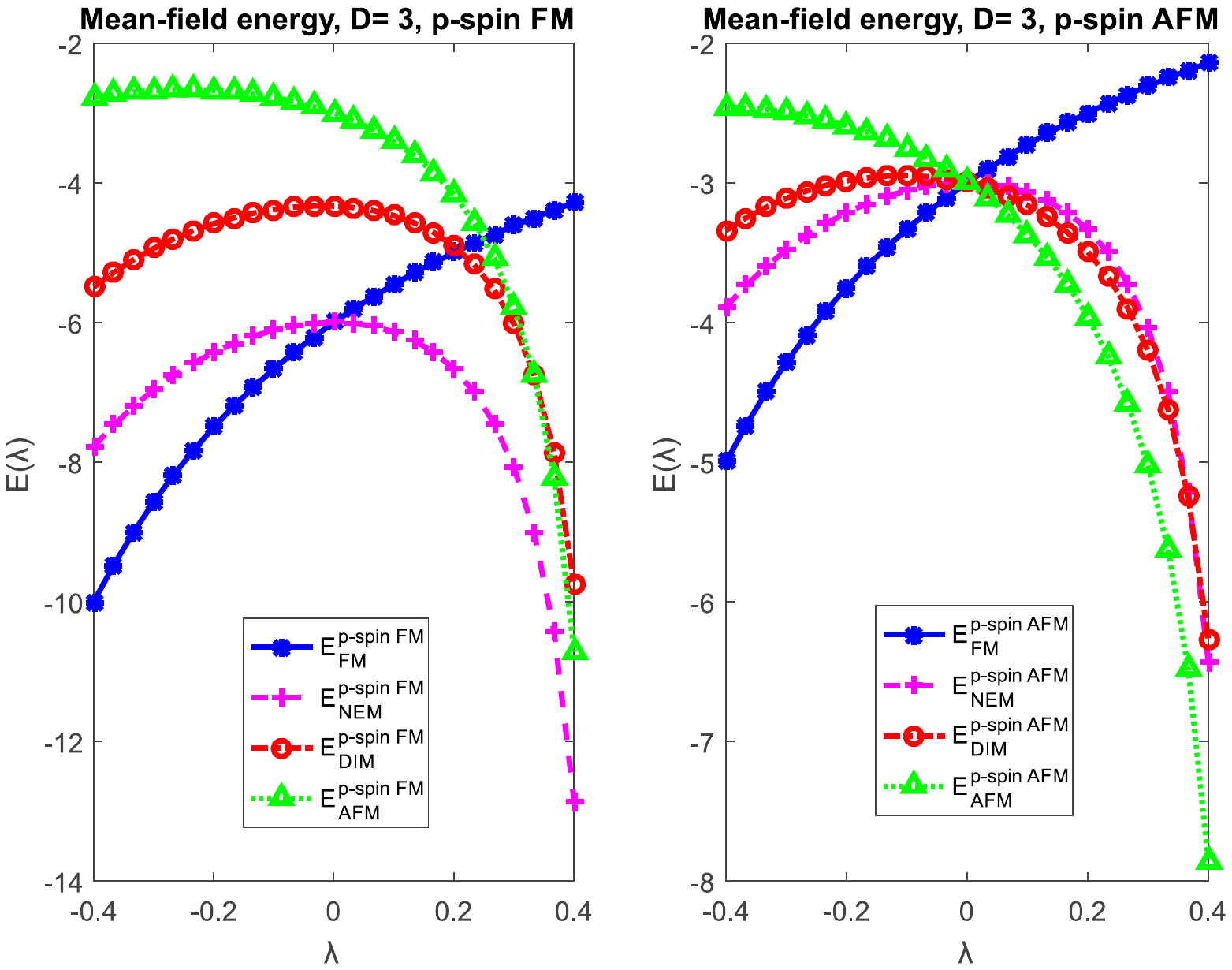}\\
  \caption{(Color online)  Mean-field energies of $H_{\eff}$ corresponding to two sorts of $S=1$ bosons on the 3D square optical lattice for the average site filling $\langle n_i \rangle= 1$ (the total number of bosons at site $i$). Energy is normalised by $E_u=2t^2/U_0$.   Left panel is for FM p-spin arrangement, right panel --- for AFM p-spin arrangement.  }\label{FigVectorBosons3D}
\end{figure}

Then we use the mean-field approach following Ref.~\onlinecite{Oles2000PRB}: we ``average''  the effective Hamiltonian over these two p-spin states, \eqref{equpup}-\eqref{equpd}. 

For FM pseudospin wave function: $\langle\Uparrow| P |\Uparrow \rangle=1$, and $\langle \Uparrow| P' | \Uparrow\rangle=0$ (see, Eqs.~\eqref{eqP}-\eqref{eqP'} for $P$ and $P'$ definitions).  
While for the second (p-AFM) case  we have: $\langle\upharpoonleft\!\downharpoonright |P|\upharpoonleft\!\downharpoonright\rangle=0$, and $\langle\upharpoonleft\!\downharpoonright|P'|\upharpoonleft\!\downharpoonright\rangle=1/2$. 

Then, for both cases $H_{\eff}$ takes the universal purely spin form:
\begin{gather}\label{eqH}
H_{\eff}\to H_{\rm spin}=\sum_{\langle i,j\rangle}\left(J_0+J_1 \mathbf S_i\cdot\mathbf S_j+J_2(\mathbf S_i\cdot\mathbf S_j)^2\right).
\end{gather}
The shadow of p-spin $\mathcal T$-space manifests itself in the exchange parameters in~\eqref{eqH}. The corresponding parameters for the FM p-spin arrangements are
\begin{align}\notag
J_0^{({\rm \scriptscriptstyle p-spinFM})}&= \frac{2  \lambda}{(1+ \lambda)(1- 2\lambda)}=\epsilon,
\\ \label{P_1}
 J_1^{({\rm \scriptscriptstyle p-spinFM})}&= - \frac{1}{1+ \lambda}=J,
 \\ \notag
J_2^{({\rm \scriptscriptstyle p-spinFM})}&= - \frac{1}{(1+ \lambda)(1- 2\lambda)}=K,
\end{align}
and for AFM  p-spin arrangements,
\begin{align}\notag
J_0^{({\rm \scriptscriptstyle  p-spin AFM})}&= -\frac{1- \lambda^2- 2 \lambda}{(1-\lambda^2)(1- 2\lambda)}=\frac{\epsilon'}{2},
\\ \label{P_2}
J_1^{({\rm \scriptscriptstyle  p-spinAFM})}&= \frac{ \lambda}{1- \lambda^2}=\frac{J'}{2},
\\ \notag
J_2^{({\rm \scriptscriptstyle p-spinAFM})}&= - \frac{\lambda^2}{(1- \lambda^2)(1- 2\lambda)}=\frac{K'}{2},
\end{align}
where the unit of energy is $E_u= 2t^2/U_0$.

In the optical lattices, spinor bosons in the Mott insulator regime can form several distinct phases, which differ in their spin correlations: ferromagnetic, antiferromagnetic, nematic (NEM), and dimer (DIM) (here, we pass by the structure of NEM and DIM phases since they are described in detail in Ref.~\onlinecite{Imambekov03}). Their energies in the mean-field approximation for the square lattice (one atom per site) are:
\begin{gather}\label{E1}
\EFM =\frac{\nu}{2}(J_0+ J_1+ J_2),
\,\,
\EAFM= \frac{\nu}{2}(J_0- J_1+ 2J_2),
\\\notag
\ENEM=\frac{\nu}{2}(J_0+ 2J_2),
\quad
\EDIM=\frac{\nu}{2} {\,} J_0 - J_1+ \frac23 J_2 (\nu+ 2),
\end{gather}
where $\nu= 2D$ is the number of nearest neighbors for the $D$-dimensional cubic lattice. Mean-field energies  as functions of $\lambda$ are shown in Figs.~\ref{FigVectorBosons} and \ref{FigVectorBosons3D} for both FM and AFM p-spin arrangements.

\begin{figure}[t]
  \centering
  \includegraphics[width=0.9\columnwidth]{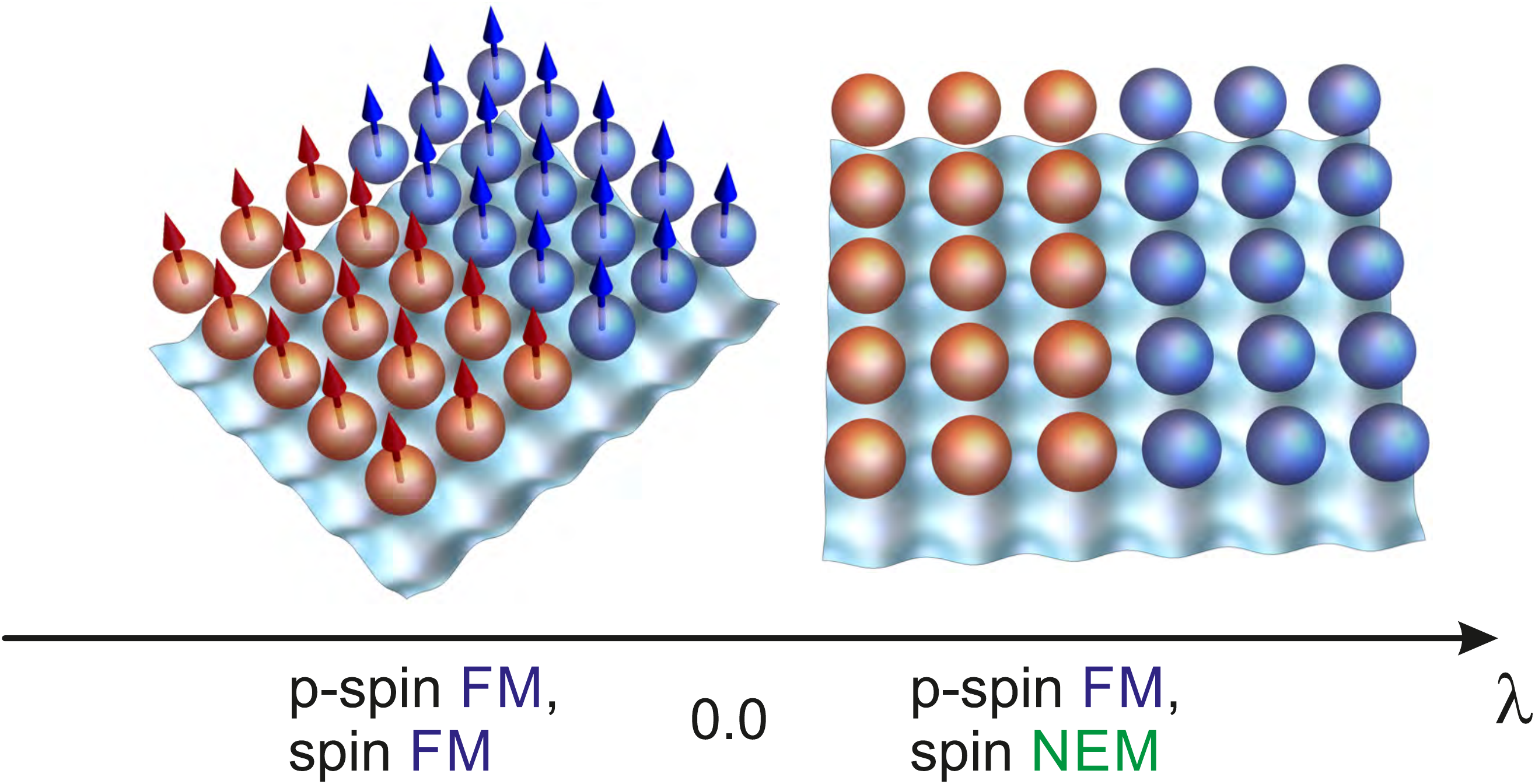}\\
  \caption{(Color online) Sketch of the phase diagram for $S=1$ bosons. We do not show spins in the spin-NEM state since its structure is rather complicated~\cite{Imambekov03}. For illustrative purposes, we schematically show spin-NEM state in Fig.~\ref{fig:spinNEM-FM} for the simple case of two sites.}\label{fig:lattices}
\end{figure}

\section{Results and Discussions~\label{sec:discuss}}

\subsection{Phase diagram for vector bosons on lattice with p-spin 1/2}
As can be seen from Figs.~\ref{FigVectorBosons} and  \ref{FigVectorBosons3D}, the  AFM p-spin state has higher energy then FM p-spin state. So, the ground state always corresponds to FM p-spin state (thus, the state shown in Fig.~\ref{lattice} is the excited one). Concentrating on the spin subsystem, we find the quantum phase transition between spin FM and spin NEM orderings at $\lambda=0$ induced by $U_s$ sign change. The transition is sketched in Fig.~\ref{fig:lattices}.
We do not show spins in the spin-NEM state since its structure is rather complicated~\cite{Imambekov03}. Note, that the p-spin FM order implies the formation of a domain structure. There arises interesting question regarding the size of domains that we leave for further investigation.

\begin{figure}[tb]
\centering
\includegraphics[width=0.99\columnwidth]{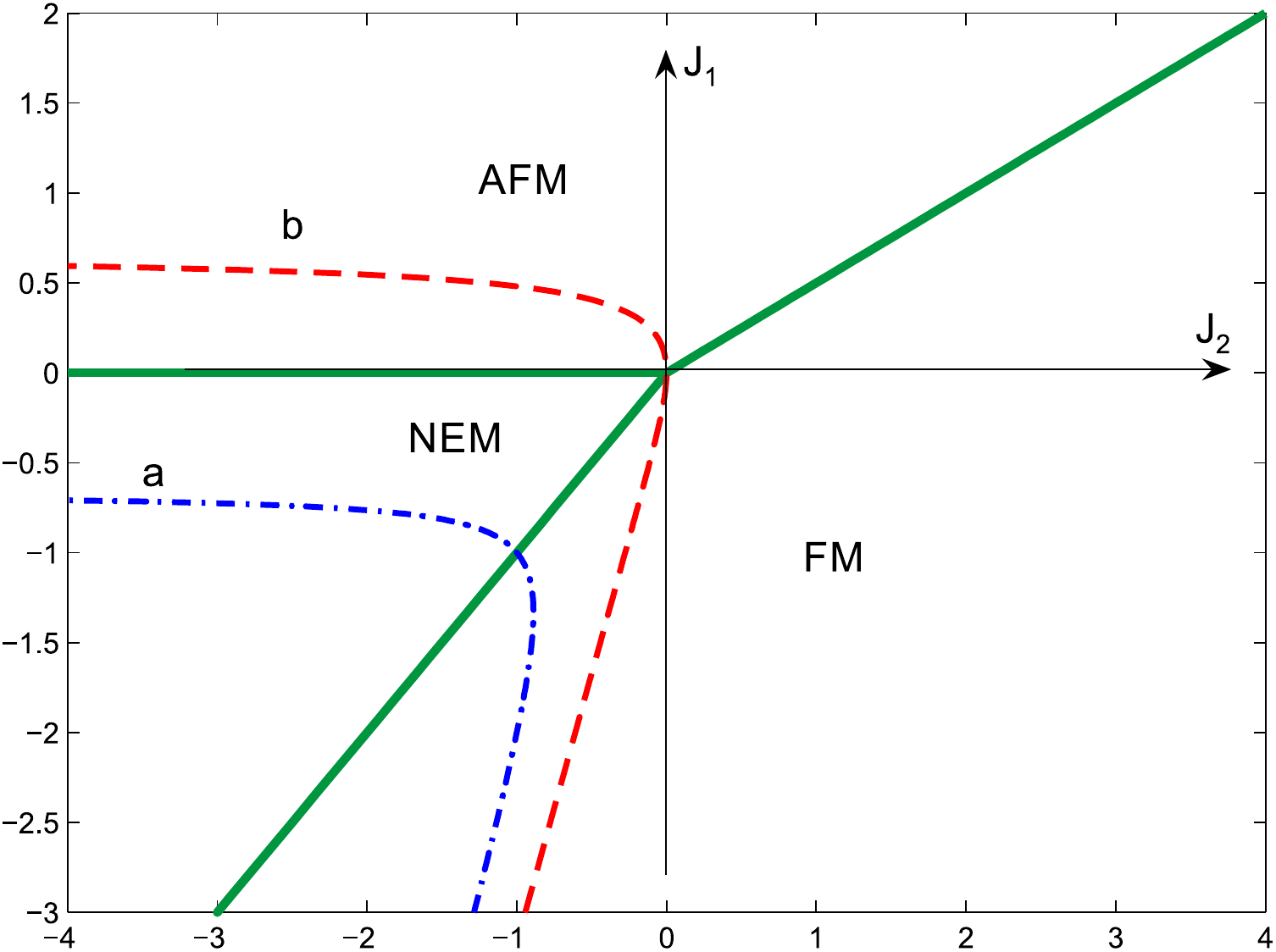}\\
\caption{\label{Fig_diag} (Color online)  Mean-field phase diagram of Hamiltonian \eqref{eqH} for arbitrary $(J_1, J_2)$ in 2D (or 3D) case. Three different spin-phases are separated by thick solid lines: ferromagnetic (FM), antiferromagnetic (AFM), and nematic (NEM). For Hamiltonian \eqref{eqH}, the  dependence of parameters $J_1$ and $J_2$ on parameter $\lambda$ ($-1 <\lambda < 1/2$) is presented by curves $a$ and $b$. Curve $a$ corresponds to the FM p-spin state and determines in the parametric form the curve ($J_1^{({\rm \scriptscriptstyle p-spinFM})}(\lambda), J_2^{({\rm \scriptscriptstyle p-spinFM})}(\lambda)$) (Eq. \eqref{P_1}), curve $b$ corresponds to the AFM p-spin state and determines in parametric form the curve ($J_1^{({\rm \scriptscriptstyle  p-spin AFM})}(\lambda), J_2^{({\rm \scriptscriptstyle  p-spin AFM})}(\lambda)$) (Eq. \eqref{P_2}). All energies are expressed in the units of $E_u$.}
\end{figure}

The above mentioned quantum phase transition can also be seen on general phase diagram of Hamiltonian \eqref{eqH}.  For arbitrary $(J_1, J_2)$ the phase diagram in 2D (or 3D) case for the average lattice site filling $n=1$ is shown in Fig.~\ref{Fig_diag}. Three different spin phases (divided by thick solid lines) are shown: ferromagnetic (FM), antiferromagnetic (AFM), and nematic (NEM)~\cite{Imambekov03}. We remind that the nematic state has zero expectation
value of any component of the on-site spin, but spin symmetry is broken since $\langle (S_i^{x,y})^2=1/2$ and  $\langle (S_i^{z})^2=0$~\cite{Imambekov03}.

For Hamiltonian \eqref{eqH}, the  dependence of parameters $J_1$ and $J_2$ on parameter $\lambda$ ($-1 <\lambda < 1/2$) is presented by curves $a$ and $b$. Curve $a$ corresponds to the FM p-spin and determines in the parametric form the curve ($J_1^{({\rm \scriptscriptstyle p-spinFM})}(\lambda), J_2^{({\rm \scriptscriptstyle p-spinFM})}(\lambda)$) (Eq. \eqref{P_1}), curve $b$ corresponds to the  AFM p-spin and defines in parametric form the curve ($J_1^{({\rm \scriptscriptstyle  p-spinAFM})}, J_2^{({\rm \scriptscriptstyle  p-spinAFM})}(\lambda)$) (Eq. \eqref{P_2}).  With increasing $\lambda$, there is one phase transition from NEM state to FM state for the curve $a$ and one phase transition from  AFM to FM state for the curve $b$. Since the curve $b$ corresponds to higher energies than the curve $a$, so $b$ describes transitions between metastable phases. The transition in the $a$ line corresponds to ground states shown  in Figs.~\ref{FigVectorBosons} and  \ref{FigVectorBosons3D}. All energies are expressed in  the units of $E_u$.

\subsection{Phase diagram for lattice fermions with p-spin~1/2}
We investigated the vector boson lattice model with p-spin 1/2. Below, we revisit the similar family of fermion models and highlight that these models give in some sense the opposite picture of quantum phase transitions.

\begin{figure}[tb]
\includegraphics[width=0.99\columnwidth]{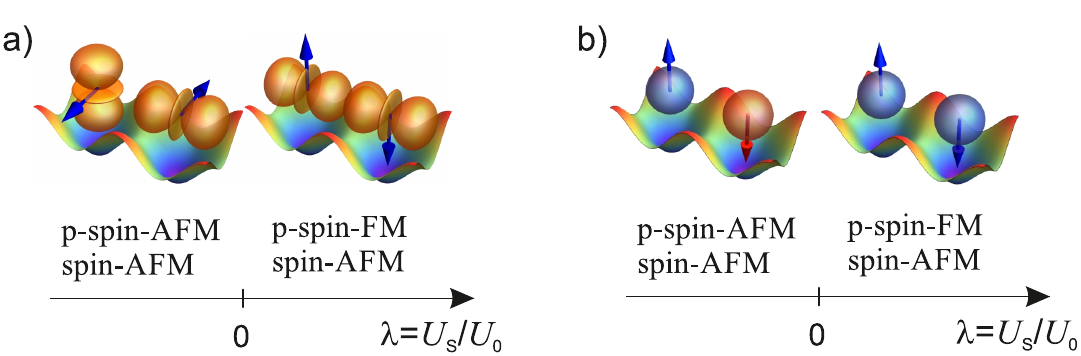}
\caption{\label{FigKH} (Color online) Contrary to two-species vector Bosons on the lattice, spin-1/2 fermions with the p-spin degree of freedom show the change of p-spin state with $\lambda$ crossing zero. (a) Standard realization of the orbital--spin model for spin-1/2 fermions with the orbital degree represented by ``real'' atomic orbitals~\cite{Kugel1982UFN}. (b) Two-species realization of the spin-1/2 fermion Kugel--Khomskii model.}
\end{figure}

We bring back the well-known Mott insulating state of fermions with spin-$1/2$ on the lattice, where fermion at each site has in addition to spin  another  degree of freedom, the ``orbital'' one, described by the quantum numbers $\alpha=1,2$, see Fig.~\ref{FigKH}. We rewrite standard results for the ground states in the notations of the present paper and compare QPTs with the boson case considered above.

This orbital degree of freedom in strongly correlated electron systems ($d$-electron compounds) corresponds to different choice of electron orbitals ai each site~\cite{Kugel1982UFN,Yamashita1998PRB}. For cold atoms on the optical lattice~\cite{li2013Nature} with two species, this case also applies, but there are other realizations, e.g., when an atom has a dipole moment (then $\alpha$ is its projection), or when the lattice site consists of two subwells (then $\alpha$ labels the atomic positions in the subwells)~\cite{Bruder2011PRA,Xu15_PRA,Chtchelkatchev2014PRA,DeChi11_PRB,
Gorshk10_NP,Porras04_PRL,Sun12_PRB,Sun14_PRB}. In most of these realizations, the interaction part of the Hubbard-like Hamiltonian describing this fermion system can be divided into two parts like for bosons above. The first term has a trivial structure in the $\alpha$-space (in our terms, p-spin space), and describes the Coulomb repulsion of fermions at one node of the lattice: $\frac12U_0\sum_{i,\sigma,\sigma',\alpha,\alpha'} n_{i\alpha\sigma} n_{i,\alpha',\sigma'}(1-\delta_{\alpha\alpha'}\delta_{\sigma\sigma'})$. The second term usually  can be expressed like the Hund's rule correlation energy~\cite{Kugel1982UFN}, $-U_s\sum_{i,\sigma,\sigma'}c_{i,1,\sigma}^\dag c_{i, 1,\sigma'} c_{i,2,\sigma'}^\dag c_{i,2,\sigma}$.

\begin{widetext}
The standard perturbative procedure with respect to the hopping amplitudes reduces again the initial model to  the Kugel--Khomskii effective  Hamiltonian~\cite{kugel1973JETP, Kugel1982UFN,Yamashita1998PRB}:
\begin{gather}\label{eqHKK}
\HKH=
J_0\sum_{\langle ij\rangle}\left\{\left(\frac14 +\textbf{S}_i\cdot\textbf{S}_j\right)[J_1+J_2\bm{{\mathcal T}}_i\cdot\bm{\mathcal T}_j+J_3{\mathcal T}_i^z{\mathcal T}_j^z]+
J_4\bm{{\mathcal T}}_i\cdot\bm{\mathcal T}_j+J_5{\mathcal T}_i^z{\mathcal T}_j^z\right\},
\end{gather}
where $\mathbf S_i$ is the fermion spin at the lattice site $i$ and $\bm{\mathcal T}_i$ is the p-spin-$1/2$ operator describing ``orbital'' degree of freedom~\cite{Yamashita1998PRB} like in Eq.~\eqref{eqSc}.
\end{widetext}
Exchange integrals $J_a$, $a=0,\ldots,5$ of  the Kugel--Khomskii Hamiltonian have been found in Ref.~\onlinecite{kugel1973JETP}. Looking at the explicit expressions for exchange integrals $J_a$ written in Eq.~6 of  Ref.~\onlinecite{kugel1973JETP} and in Ref.~\onlinecite{Yamashita1998PRB}, we see that
\begin{gather}
J_0 =\frac{4 t^2}{U_0},\qquad J_1=\frac{(1-\lambda^2-\lambda)}{2(1-\lambda^2)},
\\
 J_2=\frac{J_4}{2\lambda}=\frac{1}{1-\lambda^2},\qquad J_3=-\frac{J_5}2=\frac{2\lambda}{1+\lambda},
\end{gather}
where,  like above, $\lambda= U_s/U_0$ and the unit of energy is $E_u= 2t^2/U_0$.
So, the spin--spin and p-spin--p-spin exchange interactions change sign with $\lambda$. This behavior is the signature of a (quantum) phase transition where spin and p-spin structures of  the Mott insulating phase switch between ``ferro'' to ``antiferromagnetic'' phases, see Fig.~\ref{FigKH}. We prove it below.
\begin{figure}[t]
  \centering
  \includegraphics[width=0.45\columnwidth]{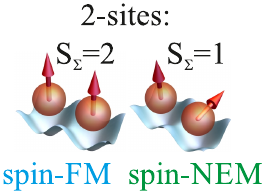}
  \caption{(Color online) The sketch illustrating the difference between the spin-NEM and spin-FM states for the trap consisting of two sites with vector bosons. }\label{fig:spinNEM-FM}
\end{figure}
\begin{figure}[bt]
  \centering
  \includegraphics[width=0.8\columnwidth]{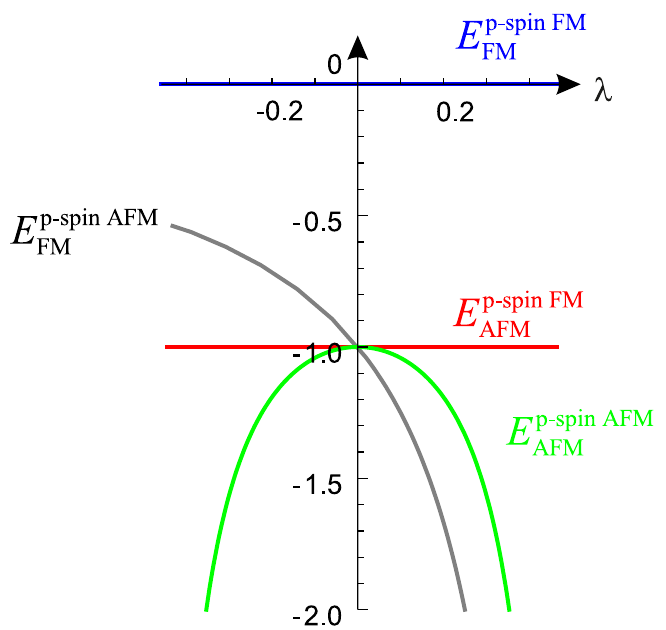}\\
  \caption{(Color online) Mean-field energies of the Kugel--Khomskii Hamiltonian $\HKH$. The energy unit is $E_u= 2t^2/U_0$.}\label{figKugelHomsky2}
\end{figure}

Here, we revisit the mean-field ground state energy of the Kugel--Khomskii Hamiltonian $\HKH$~\cite{kugel1973JETP, Kugel1982UFN,Yamashita1998PRB}. Let us consider first the FM p-spin arrangement. Then $\bm{{\mathcal T}}_i\cdot\bm{\mathcal T}_j={\mathcal T}_i^z{\mathcal T}_j^z = 1/4$ and $\HKH$  reduces to $H=J_0\sum_{\langle ij\rangle}\{\textbf{S}_i\cdot\textbf{S}_j-1/4\}$. The energies of p-spin FM and p-spin AFM states are
\begin{gather}\label{eqei}
E_{\mathrm{\scriptscriptstyle{FM}}}^{\rm p-spin\, \mathrm{\scriptscriptstyle{FM}}} = 0,
\\
E_{\mathrm{\scriptscriptstyle{AFM}}}^{\rm p-spin\, \mathrm{\scriptscriptstyle{FM}}} = -\nu E_u/2,
\end{gather}
where, like above,  $\nu= 2D$.

Let us consider now the AFM p-spin arrangement. Then  $\bm{{\mathcal T}}_i\cdot\bm{\mathcal T}_j={\mathcal T}_i^z{\mathcal T}_j^z = -1/4$ and $\HKH$ reduces to  $H=-J_0\sum_{\langle ij\rangle}\{J_4\textbf{S}_i\cdot\textbf{S}_j+\frac{J_2+J_4}4\}$. The energies of spin FM and spin AFM states are
\begin{gather}
  E_{\mathrm{\scriptscriptstyle{FM}}}^{\rm p-spin\, \mathrm{\scriptscriptstyle{AFM}}} =-\frac\nu 2E_u\frac1{1-\lambda},
   \\\label{eqef}
  E_{\mathrm{\scriptscriptstyle{AFM}}}^{\rm p-spin\, \mathrm{\scriptscriptstyle{AFM}}} = -\frac\nu 2E_u\frac1{1-\lambda^2}.
\end{gather}

The ground state energy corresponds to the minimum of these four  energy states~\eqref{eqei}-\eqref{eqef}. We illustrate this in Fig.~\ref{figKugelHomsky2} and sketch the phase diagram in Fig.~\ref{FigKH}. The ground state strongly differs from that in the cold-boson system, see Figs.~\ref{fig:lattices} and \ref{fig:spinNEM-FM}: quantum phase transition at $\lambda=0$ now changes p-spin state, see Fig.~\ref{FigKH}. That is, we see the ``inverse picture'' of quantum phase transitions compared to the boson case (where at $\lambda=0$, the spin state changes).

\section{Conclusions}
To conclude,  a generalization of the Kugel--Khomskii type model is considered, where, in contrast to the usual model, we investigate the case with spin $S=1$ and p-spin $1/2$.

We show that this model can be realized in solid state systems and also for vector two-species bosons on optical lattices.

We find the mean-field solution and discuss possible quantum phase transitions.

Finally, we revisit spin (p-spin)-1/2 Kugel--Khomskii model and see the inverse picture of phase transitions.

\begin{acknowledgments}
A.M. and A.B. are grateful to Wu-Ming Liu for interest to this work and hospitality in Beijing National Laboratory for Condensed Matter Physics. N.C. is grateful to A. Petkovic for stimulating discussions at the initial stage of this work,  to Laboratoire de Physique Th\'eorique, Toulouse, where this work was initiated, for the hospitality, and to CNRS. The research of N.M.C. was made possible by Government of the Russian Federation (Agreement № 05.Y09.21.0018).

We express our gratitude to the Computational Centers of Russian Academy of Sciences and National Research Center Kurchatov Institute for providing the access to URAL, JSCC, and HPC facilities.  This work was supported by the Russian Foundation for Basic Research (projects Nos. 16-02-00295, 16-02-00304, 17-02-00135, 17-52-53014, and 17-02-00323).
\end{acknowledgments}


\bibliography{mybibfile}

\end{document}